\documentclass{aa}

\usepackage{natbib}
\bibpunct{(}{)}{;}{a}{}{,} 
\usepackage{color}
\usepackage{graphicx}

\begin{document}

\renewcommand{\vec}[1]{\bmath{#1}}
\newcommand{\sign}{\textrm{sign}}
\newcommand{\pd}[2]{\frac{\partial #1}{\partial #2}}
\newcommand{\DS}{\displaystyle}
\newcommand{\HALF}{\frac{1}{2}}
\newcommand{\mathi}{\rm i}

\renewcommand{\vec}[1]{\mathbf{#1}}
\newcommand{\hvec}[1]{\hat{\mathbf{#1}}}
\newcommand{\av}[1]{\left<#1\right>}
\newcommand{\red}[1]{\color{red} #1 \color{black}}
\newcommand{\blue}[1]{\color{blue} #1 \color{black}}

\title{The different flavors of extragalactic jets: \\ the role of relativistic flow deceleration}

\author{}
\author{P. Rossi\inst{1}, G. Bodo\inst{1}, S. Massaglia\inst{2}, A. Capetti\inst{1} }

\authorrunning{P.Rossi et al.}
\titlerunning{Deceleration of relativistic jets}

\institute{INAF/Osservatorio Astrofisico di Torino, via Osservatorio 20, 10025 Pino Torinese, Italy  \and Dipartimento di Fisica, Universit\`a degli Studi di Torino, via
  Pietro Giuria 1, 10125 Torino, Italy}

\date{Received ?? / Accepted ??}


\label{firstpage}

\abstract
{We perform three-dimensional numerical simulations of relativistic (with a Lorentz factor of 10), non magnetized jets propagating in a uniform density environment, in order to study the effect of the entrainment and the consequent deceleration. Our simulations investigate the jet propagation inside the galaxy core, where, most likely,  the deceleration occurs more efficiently. We compare cases with different density and pressure ratios with respect to the ambient medium finding that low density jets are efficiently decelerated and reach a quasi steady state in which, over a length of 600 jet radii, slow down from highly to sub-relativistic velocities.
At the opposite, denser jets keep highly relativistic velocity over the same length. We discuss these results in relation to the Faranoff Riley (FR) radio-sources classification. We infer that lower density jets can give rise to FR 0 and FR I radio-sources, while higher density jets may be connected to FR II radio-sources.
}

\maketitle

\section{Introduction}
%
%
%
 
 Extragalactic radio sources are traditionally divided into two morphological classes
according to their intrinsic radiative power \citep{FR74}: low luminosity sources
(Fanaroff-Riley type I, FR I) are brighter close to the nucleus of the parent galaxy and
their jets become dimmer with distance, while high power sources (Fanaroff-Riley type
II, FR II) show the maximum brightness in the hot spots at the jet termination. The
different morphology is generally accepted as reflecting a difference in how the jet
energy is dissipated during propagation in the ambient medium, and produces
the observed radiation. In FR II sources, energy and momentum are transported
without strong losses to the jet termination, while in FR I sources, turbulence and
entrainment must play an important role in shaping their morphology. Recently, a third class has been introduced by \citet{baldi15} and referred to as FR 0. The sources belonging to this type are highly core dominated and generally do not present the prominent extended radio structures typical of FR I and FR II. Nonetheless, evidence for small scale jets (limited to sizes of a few kpc) are found in a minority of FR 0 \citep{baldi19}). This suggests that their jets are not even able to escape from the galaxy core.

Observational evidences show that, both in FR II and in FR I, the jets at their base (at
the parsec scale) are relativistic, with very similar Lorentz factors \citep{gg01,
celotti08}. \citet{massaro20} found that the FR 0 represents the misaligned counterparts of a significant fraction of BL~Lac objects: BL~Lacs are known to be associated with highly relativistic jets \citep{blandford78} and this result implies that at least some FR 0 must also launch relativistic outflows. It is then clear that in FR I jets a deceleration to sub-relativistic velocities must occur between the inner region and the kiloparsec scale, where they show their typical plume-like, turbulent morphology \citep{laing02,laing14}. The same effect must occur also in FR 0, at even smaller distances from the jet origin. Assuming that the deceleration to a sub-relativistic velocity occurred inside the galaxy core, \citet{massaglia16} (hereinafter Mas16 ) performed
high-resolution three-dimensional Newtonian simulations of low Mach number jets, showing how
turbulence develops and gives rise to a jet structure very similar to that observed in
FR I sources.  Recently, \citet{westhuizen19} using injection parameters similar to those of  Mas16 , and applying a recipe for synchrotron radiation model investigated whether the emission map could be comparable to FR I observations.  

A very likely mechanism, through which deceleration occurs, is entrainment of material from the ambient medium \citep{DeYoung93, Bicknell84, Bicknell86, Bicknell94, DeYoung05}, promoted by different kinds of jet instabilities \citep[see, e.g.,][]{perucho05, peru07, meliani07, meliani09,  perucho10, bodo13, matsumoto13, millas17, toma17, komissarov18, Tchekhovskoy16, mukherjee20}. An other possibility suggested by \citet{Komissarov94} is mass loading by stellar winds; subsequently this approach has been investigated by
\citet{Bowman96, hubbard06, perucho14}  and more recently \citet{ perucho20} suggests that stars may trigger mixing by generating small scale instabilities.

Since FR I jets are less powerful than FR II, but have similar Lorentz factors, they have
to be less dense and therefore more prone to deceleration by the external medium.
\citet{rossi08} performed numerical simulations of the propagation of relativistic
jets with different values of the density ratio between the jet and the ambient medium
and showed that, while jets with density ratio between $0.01-0.1$ (and power
corresponding to FR II) propagate almost undisturbed, jets with lower values of the
density ratio (and power corresponding to FR I) show evidences of entrainment and
deceleration in their external layers.  Due to the limitations of computational power,
\citet{rossi08} could follow the jet propagation only up to about $60$ jet radii  and could not see a complete transition to sub-relativistic velocities, in fact in their simulations the jet core remained relativistic.

In this paper we extend the work by \citet{rossi08} by following the jet propagation up to $600$ jet radii, a distance that, assuming the initial jet radius of the order of 1 pc, covers the full galaxy core. We consider similar jet parameters and the corresponding jet powers are at or below FR I -- FR II transition value. In Section \ref{sec:numsetup} we describe the set of equation to be solved, the numerical method, the initial and boundary conditions and the parameters of the simulations. In Section \ref{sec:results} we present our results and in Section \ref{sec:discussion} we give a discussion and conclusions.

\section{Numerical setup}
\label{sec:numsetup}
%
%
%
%
%
%
\begin{table*}[!ht] 
\centering
\begin{tabular}{crrrrcccc}\hline                                        
Case    & $\eta$      & $K$    &$M$    &$ M_r$ & $L_x\times L_y\times L_z$   & $N_x\times N_y\times N_z$ & $P_{\mathrm j} \ (\mathrm {erg \ s}^{-1})$  \\
\hline\hline\noalign{\medskip}
A      & $10^{-3}$        & $1$     & $3$   &$28$ & $240 \times 600\times 240$  & $500\times2600\times500$ & $1.5 \times 10^{44}$  \\
B      & $10^{-4}$        & $1$     & $3$   &$28$ & $450 \times  600\times 450$  & $570\times2600\times570$  & $1.5 \times 10^{43}$  \\
C      & $10^{-4}$        & $10$   & $1.8$   &$15$ & $300 \times  600\times 300$  & $520\times2600\times520$  & $4.2 \times 10^{43}$  \\  
D      & $10^{-2}$        & $1$   & $3$   &$28$ & $120 \times  600\times 120$  & $360\times 2600\times 360$  & $1.5 \times 10^{45}$  \\   
 \noalign{\medskip}
\hline
\medskip
\label{tab:cases} 
\end{tabular}

\caption[]{Parameter set used in the numerical simulation model, the second 
column refers to density ratio $\eta$, the third to the  pressure ratio $K$, the fourth to the classical Mach number, the fifth to the relativistic Mach number, the sixth to the size of the physical domain in unit of the jet radius and the seventh the  number of grid points and the last one the jet power evaluated according to equation (\ref{eq:jetpower})}
\label{parset}
\end{table*}

Numerical simulations are carried out by solving 
the equations of particle number and energy-momentum conservation.
Referring to the observer's reference frame, where the fluid moves with 
velocity $v_k$ (in units of the speed of light $c$) with respect to the coordinate 
axes $k=x,y,z$ and assuming a flat metric, the 
conservation laws take the differential form:  
\begin{equation}\label{eq:conslaw}
 \pd{}{t}\left(\begin{array}{c}
  \rho\gamma \\ \noalign{\medskip}
  w \gamma^2 v_k  \\ \noalign{\medskip}
  w \gamma^2 - p  \\ \noalign{\medskip} 
  \rho\gamma f  
\end{array}\right) 
  +
 \sum_i\pd{}{x_i} \left(\begin{array}{c}
 \rho\gamma v_i                   \\ \noalign{\medskip}
 w\gamma^2 v_k v_i + p\delta_{ki} \\ \noalign{\medskip}
 w \gamma^2 v_i                   \\ \noalign{\medskip}
 \rho\gamma f v_i
\end{array}\right) = 0 \,,
\end{equation}
where $\rho, w, p$ and $\gamma$ denote the rest 
mass density, enthalpy, gas pressure and Lorentz factor, respectively.
The jet and external material are distinguished by using a passive tracer, $f$, 
set equal to unity for the injected jet material and equal to zero for the 
ambient medium.
We take the jet radius as the unit of length, the speed of light as the unit of velocity and the light crossing time of the jet radius as the unit of time. The system of equations (\ref{eq:conslaw}) 
is completed by specifying an equation of state relating $w$, $\rho$ and $p$. 
Following \citet{Mignone05}, we adopt the Taub-Matthews (TM) equation of state with the  prescription: 
\begin{equation}
 w = \frac{5}{2}p + \sqrt{\frac{9}{4}p^2 + \rho^2}
\end{equation}
which closely reproduces the thermodynamics of the Synge gas for a 
single-species relativistic perfect fluid.

Simulations are carried out on a Cartesian domain with coordinates 
in the range $x\in [-L_x/2,L_x/2]$, $y\in [0,L_y]$ and $z\in [-L_z/2,L_z/2]$
(lengths are expressed in units of the jet radius).
At $t=0$, the domain is filled with a perfect fluid of uniform density 
and pressure, representing the external ambient medium initially at rest. The assumption of constant external density is consistent with the fact that we focus on the jet propagation inside the galaxy core.
 
A cylindrical inflow of velocity $v_j$ is  constantly fed into the domain 
through the lower $y$ boundary along $y$ direction.
The jet is characterized by the Lorentz factor $\gamma_j$,  by the ratio $K$ between the proper  ($p_j$) and 
 the ambient ($p_0$) pressures  and the ratio  $\eta$ between the proper ($\rho_j$)  and the ambient ($\rho_0$) densities.
 The jet Mach number can be derived using the expression of the sound speed for the TM equation of state  \citep{Mignone05}:
\begin{equation}
 c_s^2=\frac{p}{3w} \frac{5w-8p}{w-p}
\end{equation}
The classical Mach number is defined as $M= v_{\mathrm j}/c_{\mathrm s}$, 
in the relativistic case we can define a generalization \citep{konigl80} as $M_{\mathrm r}=\gamma_{\mathrm j} v_{\mathrm j}/\gamma_{\mathrm s} c_{\mathrm s}$, where $\gamma_{\mathrm s}=1/\sqrt{1-c_{\mathrm s}^2}$.
Outside the inlet region we impose symmetric boundary conditions (emulating a counter-jet), 
whereas the flow can freely leave the domain throughout the remaining boundaries. 

Explicit numerical integration of the equations (\ref{eq:conslaw}) is achieved
using the relativistic hydrodynamics module available in the 
PLUTO code \citep{PLUTO}. 
For the present application, we employ linear reconstruction and HLLC Riemann solver \citep{HLLC}. 
The physical domain is covered by $N_x\times N_y\times N_z$ computational zones,
not necessarily uniformly spaced.
For domains with a large physical size, we employ a uniform grid resolution around 
the beam (typically for $|x|,|z| \le 7$) and a geometrically stretched grid elsewhere. 
 
 We perform a set of four simulations, whose parameters are reported in Table \ref{tab:cases}. All the cases have the Lorentz factor $\gamma = 10$, cases A and B and D differ by the density ratio $\eta$, which is, respectively $10^{-3}, 10^{-4}$ and $10^{-2}$, while cases B and C have the same density ratio and differ by the pressure ratio $K$.

\section{Results}
\label{sec:results}
%
%
\begin{figure*}[!ht] 
\centering%
\includegraphics[width=1.5\columnwidth]{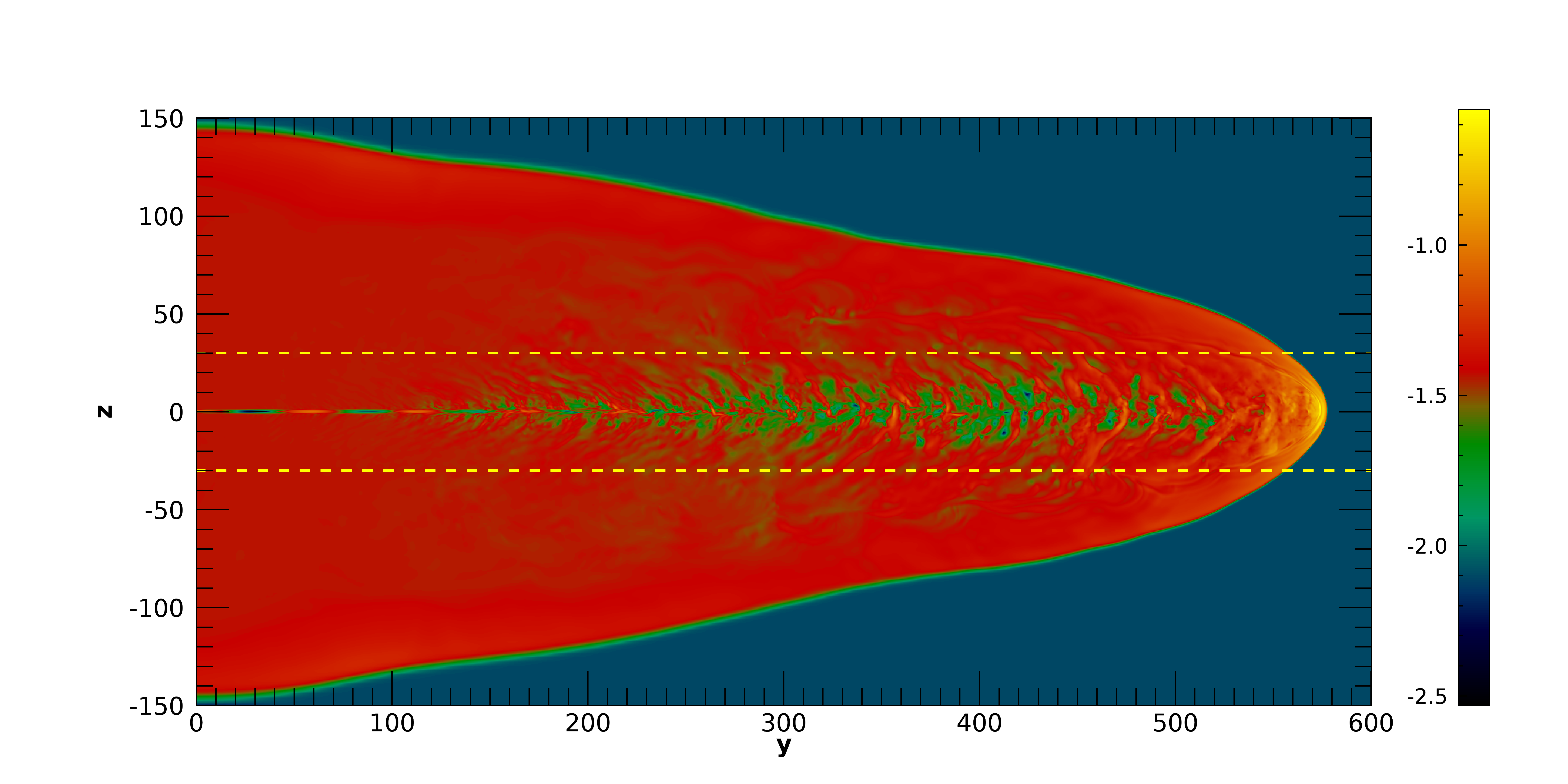}%
\caption{Longitudinal cut (in the $y-z$ plane) of the distribution of the logarithm of the pressure for case C at $t=11500$, when the bow shock in front of the jet has reached $y=580$. The image shows the whole computational domain. The two dashed yellow lines indicate the central region on which we will focus in our following analysis.  
}
\label{fig:fig1} 
\end{figure*}

\begin{figure*}[!ht] 
\centering%
\includegraphics[width=1.5\columnwidth]{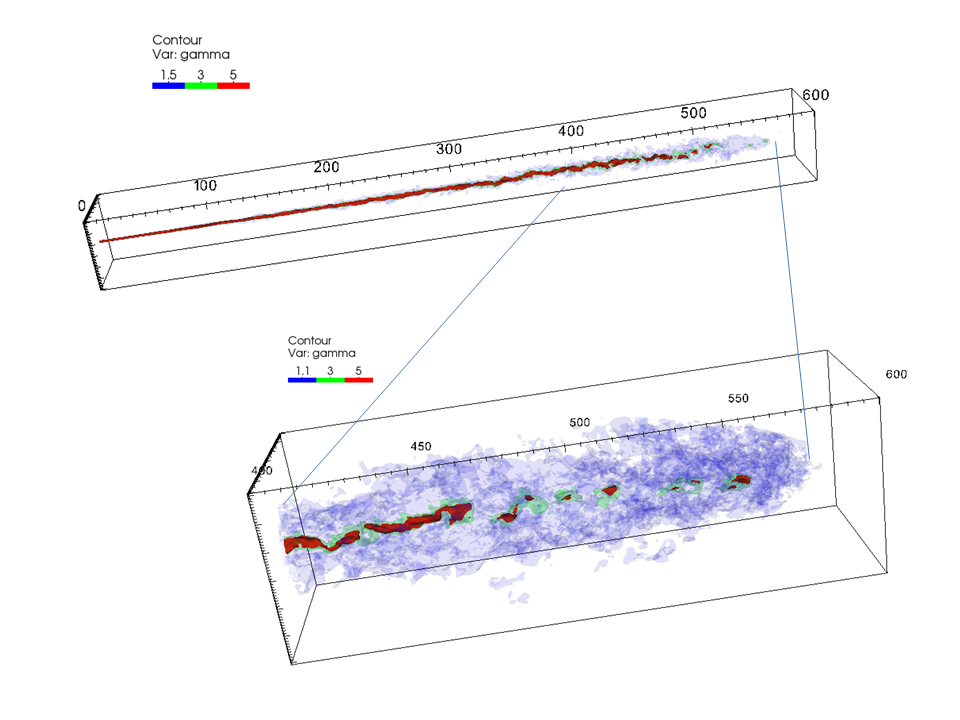}%
\caption{3D isosurfaces of the Lorentz factor for case A, the top panel shows the entire jet, while the bottom panel the last portion between $y=400$ and $y=600$. The snapshot is taken at $t=6,000$ when the jet head reaches $y\sim 600$. the three isosurfaces are for $\gamma=1.5,3,5$ in the top panel and $\gamma=1.1,3,5$ in the bottom panel
}
\label{fig:caseA} 
\end{figure*}

\begin{figure*}[!ht] 
\centering%
\includegraphics[width=1.5\columnwidth]{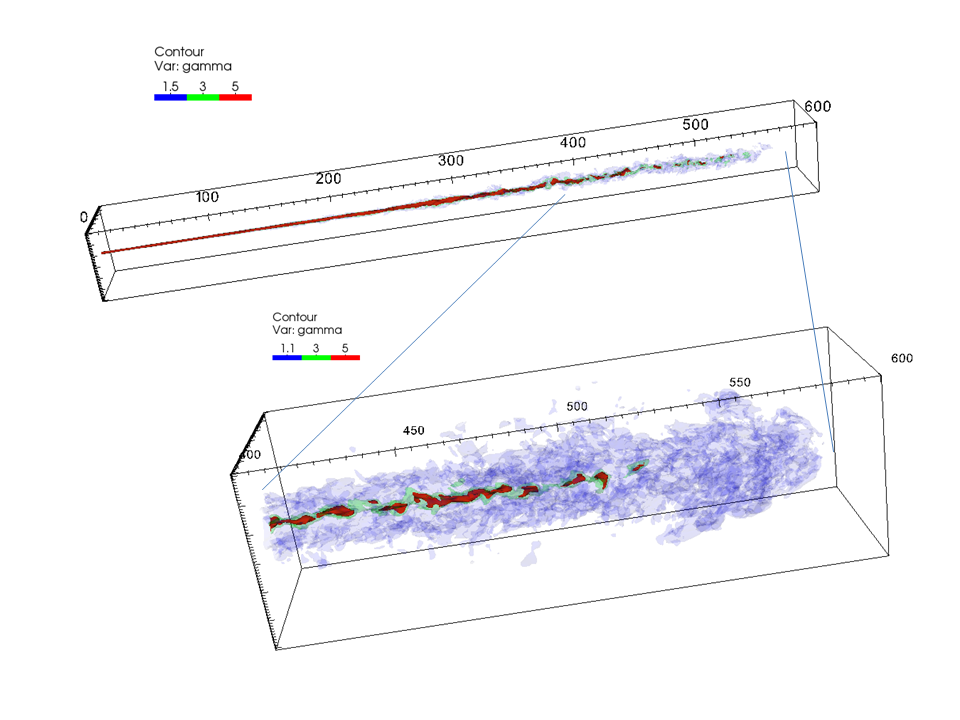}%
\caption{3D isosurfaces of the Lorentz factor for case B, the top panel shows the entire jet, while the bottom panel the last portion between $y=400$ and $y=600$. The snapshot is taken at $t=25,000$ when the jet head reaches $y\sim 600$. the three isosurfaces are for $\gamma=1.5,3,5$ in the top panel and $\gamma=1.1,3,5$ in the bottom panel
}
\label{fig:caseB} 
\end{figure*}

\begin{figure*}[!ht] 
\centering%
\includegraphics[width=1.5\columnwidth]{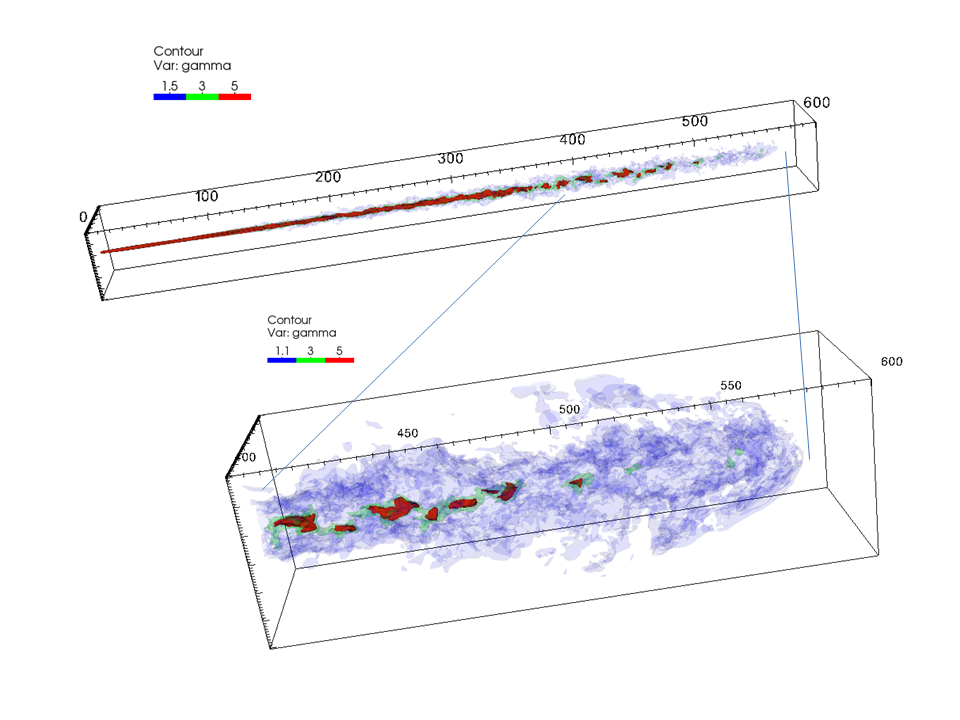}%
\caption{3D isosurfaces of the Lorentz factor for case C, the top panel shows the entire jet, while the bottom panel the last portion between $y=400$ and $y=600$. The snapshot is taken at $t=12,500$ when the jet head reaches $y\sim 600$. the three isosurfaces are for $\gamma=1.5,3,5$ in the top panel and $\gamma=1.1,3,5$ in the bottom panel.
}
\label{fig:caseC} 
\end{figure*}

Our aim is to understand how the deceleration of a relativistic jet depends on its physical parameters.  In principle, the setup of our simulations is such that the results are scale invariant, however our interest is focused on the deceleration that mainly occurs in the first kiloparsec. This motivates our assumption of constant density of the external medium, because the domain of our simulations can be assumed to be all contained within the galaxy core. We can then adopt a jet injection radius of the order of one parsec and the external density at injection of the order of one particle per cubic centimeter \citep{balm08}.  

In the following we will express all quantities in their non dimensional values as defined in the previous Section. Conversion to physical units can be obtained with the above assumptions, in particular the unit of time is:
\begin{equation}
    \tau \sim 3.25 \left(\frac{r_{\mathrm j}} {1 \mathrm{pc} } \right) \mathrm{yrs}
\end{equation}
 and the jet power is:
\begin{equation}\label{eq:jetpower}
P_{\mathrm j} \sim h \left(\frac{\gamma}{10}\right)^2 \left(\frac{\eta}{10^{-4}}\right)\left(\frac{r_{\mathrm j}} {1 \mathrm{pc} } \right)^2
\left(\frac{n_0}{1 {\mathrm {cm}}^{-3}} \right) 1.22\times 10^{43}\; {\mathrm {erg \ s}^{-1}}
\end{equation}
where $h$ is the specific enthalpy and $n_0$ is the external number density at the injection point.
 The jet  power for the four cases (with specific enthalpy $h = 1.2, \, 1.2, \, 3.45$ and $1.2$ respectively) are reported in Table \ref{tab:cases}. We choose the jet parameters so that the resulting powers span an interval between FR I and FR II. Case D can be taken as a reference case, being clearly on the FR II side.

  We follow the jet propagation until the bow shock in front reaches $y=600$, the lateral extension of the domain is such that the entire bow shock is included in the computational domain. In Fig. \ref{fig:fig1} we show a longitudinal cut (in the $y-z$ plane) of the distribution of the logarithm of the pressure for case C at $t=11500$, when the bow shock in front of the jet has reached $y=580$. The image shows the whole computational domain. The two dashed yellow lines indicate the central region on which we will focus in our following analysis.
In Figs. \ref{fig:caseA}-\ref{fig:caseC} we show the jet structure for the first three cases, when the bow shock in front of the jet reaches $600$ length units.  These three snapshots correspond to three different times, namely $t=6,000$ for case A, $t=25,000$ for case B and $t=12,500$ for case C,  the differences in times are due to the different jet head propagation velocity, as we will see below. In the figures we display three dimensional views of the isocontours for three value of Lorentz gamma factor. In the top panel of each figure we have the entire jet, while in the bottom panel we zoom on the last $200$ length unit. The represented domain, in the transverse section, spans from $-30$ to $30$  in both directions. In all the three cases we can see that there is a first region (up to $y \sim 150-250$) in which the jet seems to propagate straight, although, as we will see below perturbations are already present and growing. Then we observe perturbations that lead to a displacement of the jet axis. Finally, in the last part of the jet, we see that the fast moving material fragments (see bottom panels) and we have a wide low velocity envelope in which a few blobs still move at high $\gamma$.

 \begin{figure*}[!ht] 
\centering%
\includegraphics[width=2\columnwidth]{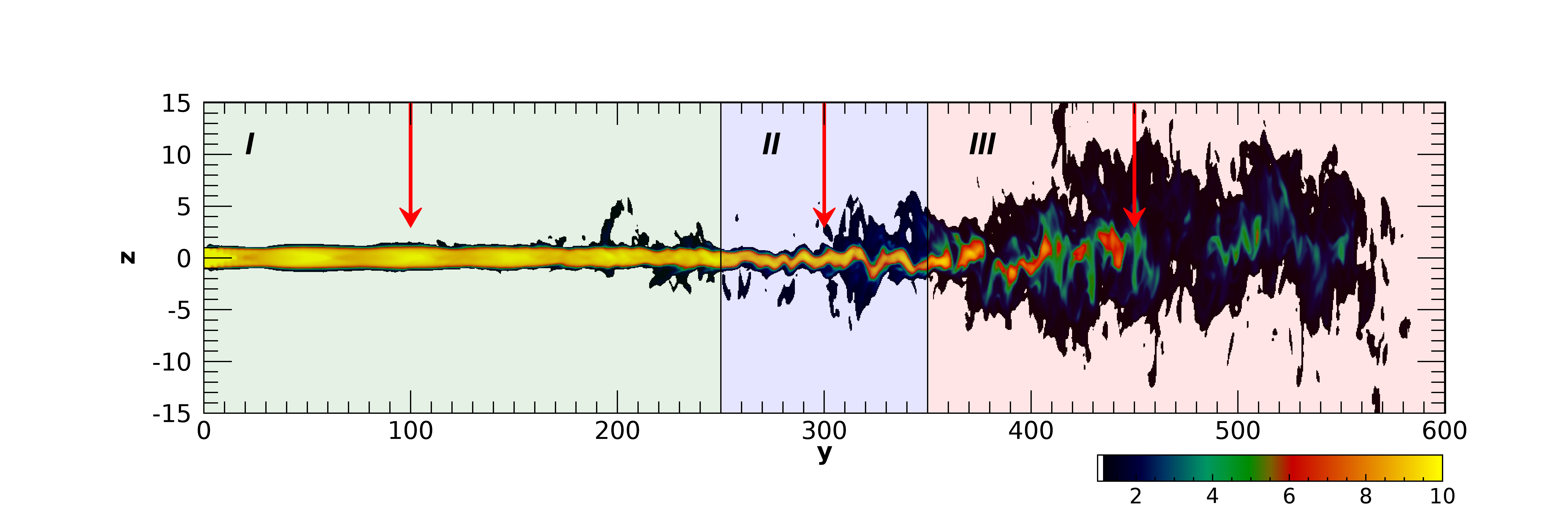}%
\caption{Longitudinal cut (in the $y-z$ plane)  of the distribution of the Lorentz factor for case B at $t=25,000$ when the jet head reaches $y=600$. The three colored bands highlight the three phases described in the text. The red arrows mark the positions at  which are taken the transverse cuts shown in Fig. \ref{fig:gammacut_caseB}.
}
\label{fig:longcut} 
\end{figure*}

\begin{figure*}[!ht] 
\centering%
\includegraphics[width=2\columnwidth]{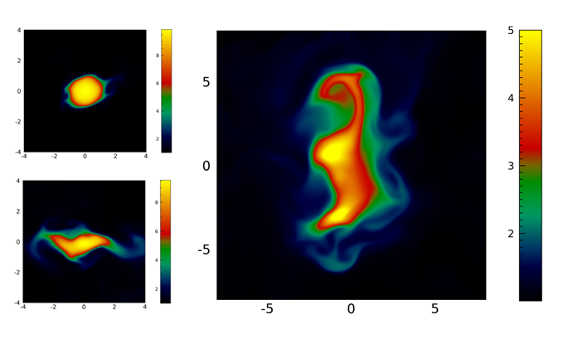}%
\caption{Transverse cuts (in the $x-z$ plane) of the distribution of the Lorentz factor at three different positions along the jet, for case B at $t=25,000$ when the jet head reaches $y=600$. The positions of the three cuts correspond to the three red arrows in Fig. \ref{fig:longcut}, the top left panel is at $y=100$, the bottom left panel is at $y=300$ and the right panel at $y=450$. Notice that the two left panel cover a square with $-4\leq x,z \leq4$, while for the right panel $-8\leq x,z \leq8$, moreover in the two left panels the maximum $\gamma$ is $10$, while in the right panel the maximum $\gamma$ is $5$. 
}
\label{fig:gammacut_caseB} 
\end{figure*}

\begin{figure*}[!ht]  
\centering%
\includegraphics[width=2\columnwidth]{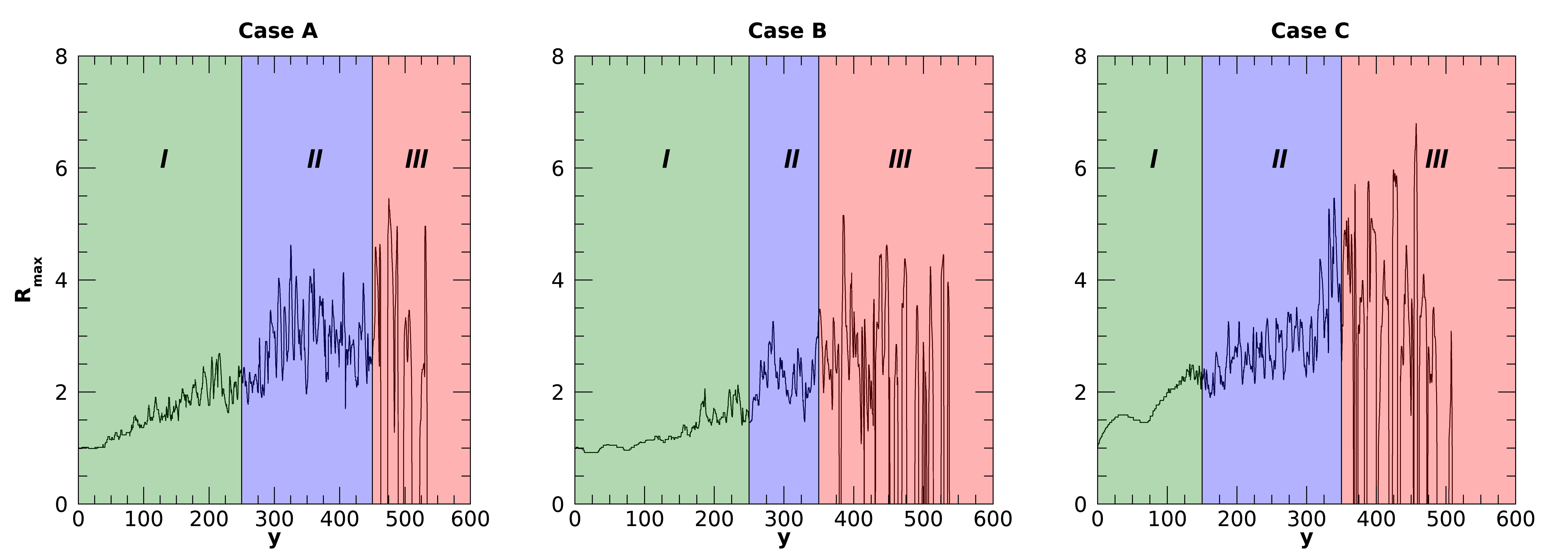}%
\caption{Plots of the maximum distance from the initial jet axis at which we find material moving at $\gamma\geq 5$  as a function of $y$. The three panels refer to the three cases, the times are respectively $t=6,000, \ 12,500, \ 25,000$ when the jet head reaches $y=600$. The three colored bands highlight the three phases described in the text.
}
\label{fig:r_max} 
\end{figure*}

In order to better understand the jet evolution, outlined above,  we will discuss in more detail case B for which in Fig. \ref{fig:longcut}  we display  a longitudinal cut (in the $y-z$ plane) of the distribution of the Lorentz factor at $t=25,000$ when the jet head reaches $y=600$.  
We highlight the three regions of different jet behavior  with different colors: in the first region (up to $y \sim 250$, region {\it I})  we see a series of recollimation shocks that modulate the jet radius. Starting from the second recollimation shock we notice the formation and growing of perturbations, leading, in region {\it II}, to a displacement of the jet from its original axis, and  to entrainment of slow moving material.  
 Finally, in the last region, the jet is broken in few fast moving blobs surrounded  by a wide low velocity envelope (region {\it III}).

In Fig. \ref{fig:gammacut_caseB} we show three transverse (in the $x-z$ plane) cuts of the distribution of the $\gamma$ factor, their positions are indicated with red arrows in Fig. \ref{fig:longcut}. The three panels show different stages of the evolution of the perturbations: at the beginning the jet cross section is slightly deformed, then becomes completely distorted and in the last panel appears to be more spread and the maximum value of the Lorentz factor is reduced to $\gamma =5$. The last $x-z$ cut is through a fast moving blob, in other position the maximum $\gamma$ is $ \approx 2$.

Also the other two cases (A and C) show a similar behavior, that we can analyze by looking at Fig. \ref{fig:r_max}, where we plot the maximum distance from the jet axis, at which we still find jet material moving with $\gamma \geq 5$. In the figure we identify the three defined regions, whose lengths and starting points  are different for the three cases, with different colors.  
For case A (left panel) we observe that perturbations start to form and  grow at $y=50$ and from $y=250$ they induce jet oscillation. At $y=450$ the jet breaks and in fact we  see intervals in which there is no jet material with $\gamma \geq 5$. In the middle panel we present case B, that  displays what we have already discussed referring to Fig. \ref{fig:longcut}. In particular we highlight the modulation of the jet radius by recollimation shocks (between $y=0$ and $y=200$); the growing of perturbations (region $II$) and the  final  jet disruption. In case C (right panel), we see a sudden expansion, due to the fact that the jet is overpressured, modulated by recollimation shocks up to $y=150$, then perturbations grow and lead to oscillation of smaller amplitude with respect to the case A. The jet breaking occurs at $y > 350$.

\begin{figure*}[!ht] 
\centering%
\includegraphics[width=2\columnwidth]{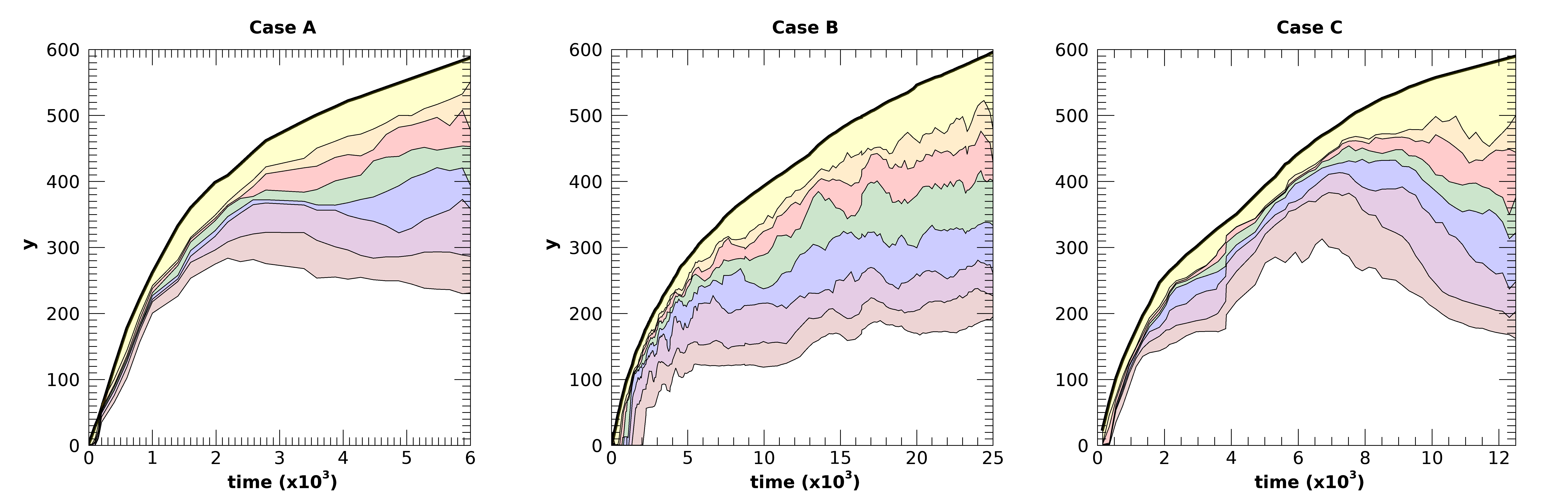}%
\caption{ The three panels show plots of the maximum distance from the jet origin at which a given fraction of momentum flux is carried by material moving at $\gamma \geq 5$ as a function of time.
From the top we show: the jet head position (black thick curve), and the curves from $10\%$ to $70\%$ differing by $10\%$. We highlight the areas between subsequent curves with different colors.
}
\label{fig:fig7.png} 
\end{figure*}

\begin{figure}[!ht]
\centering%
\includegraphics[width=\columnwidth]{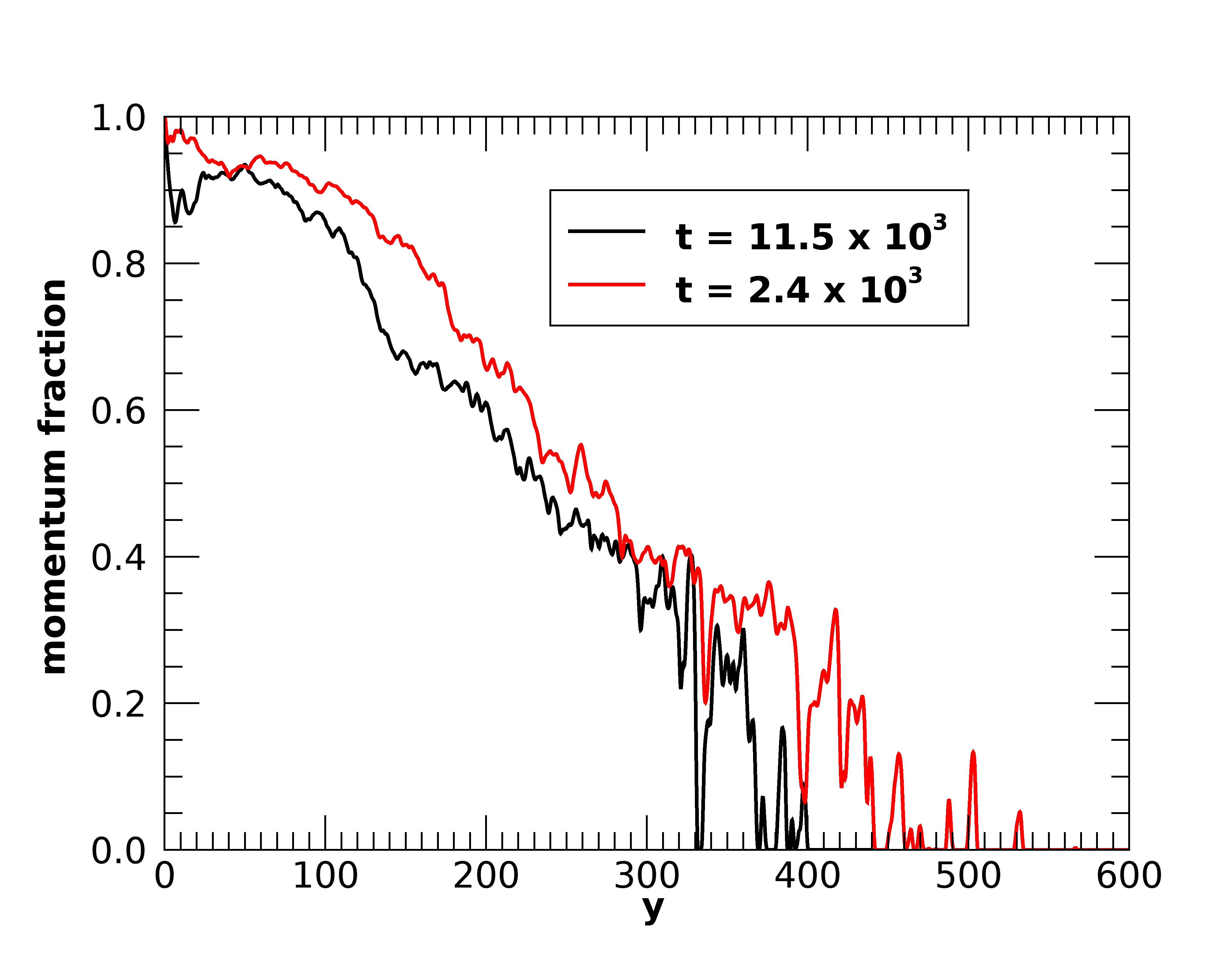}%
\caption{ Plot of the fraction of momentum flux carried by material moving at $\gamma \geq 5$ as a function of $y$ at two different times, for case B.
}
\label{fig:figmomfr} 
\end{figure}

The results presented above show that the jet progressively decelerates and the quantity of material moving at high $\gamma$ decreases, while an increasing amount of material around the jet moves at $\gamma \leq 2$. 

Now we can analyze more quantitatively how the deceleration occurs or, in other words, how the jet transfers its momentum to the external medium. In Fig. \ref{fig:fig7.png}  we plot, as a function of time, the maximum distance at which a given fraction of the momentum flux is carried by material moving at $\gamma \geq 5$. For example, looking at the left panel (case A) we have eight curves enclosing $7$ colored regions. Starting from the top, the black thick curve represents the head position, the first region   refers to a jet section in which the high velocity material carries a fraction of momentum flux $\leq 10\%$, the second band refers to a fraction $10\%-20\%$ and so on, the bottom curve refers to a fraction of $70\%$. Up to $t \approx 2,000$, all the curves follow the behavior of the jet head, and all the momentum transfer occurs close to the terminal shock. Subsequently, the  curves, starting from the bottom progressively flatten.  The flattening of the curves indicates that the jet has reached a quasi steady state up to that position, for example the $70\%$ curve (the lowest one) flattens at $y \approx 250$. This means that the fast moving material, up to that point, looses in a steady way $30\%$ of its momentum flux,  in other words, $70\%$ of the momentum flux of the fast moving material goes beyond $y \approx 250$ and this fraction remains constant in time. Between $y=250$ and $y=450$ the jet material moving at $\gamma \geq 5$ carries a fraction of momentum flux decreasing from $70\%$ to about $30\%$. The $10\%$ and $20\%$ curves still stay parallel to that of the head position. The last portion of the jet has not yet reached a steady configuration and the remaining $\sim 30\%$ of momentum is transferred in the region close to the terminal shock. Case B shows a similar behavior, however the flattening of the curves starts at an earlier time, moreover also the $20\%$ and $10\%$ curves show a more pronounced change of slope. Case C has a non-monotonic behavior and will be discussed in more detail below. 

\begin{figure*}[!ht] 
\centering%
\includegraphics[width=6cm]{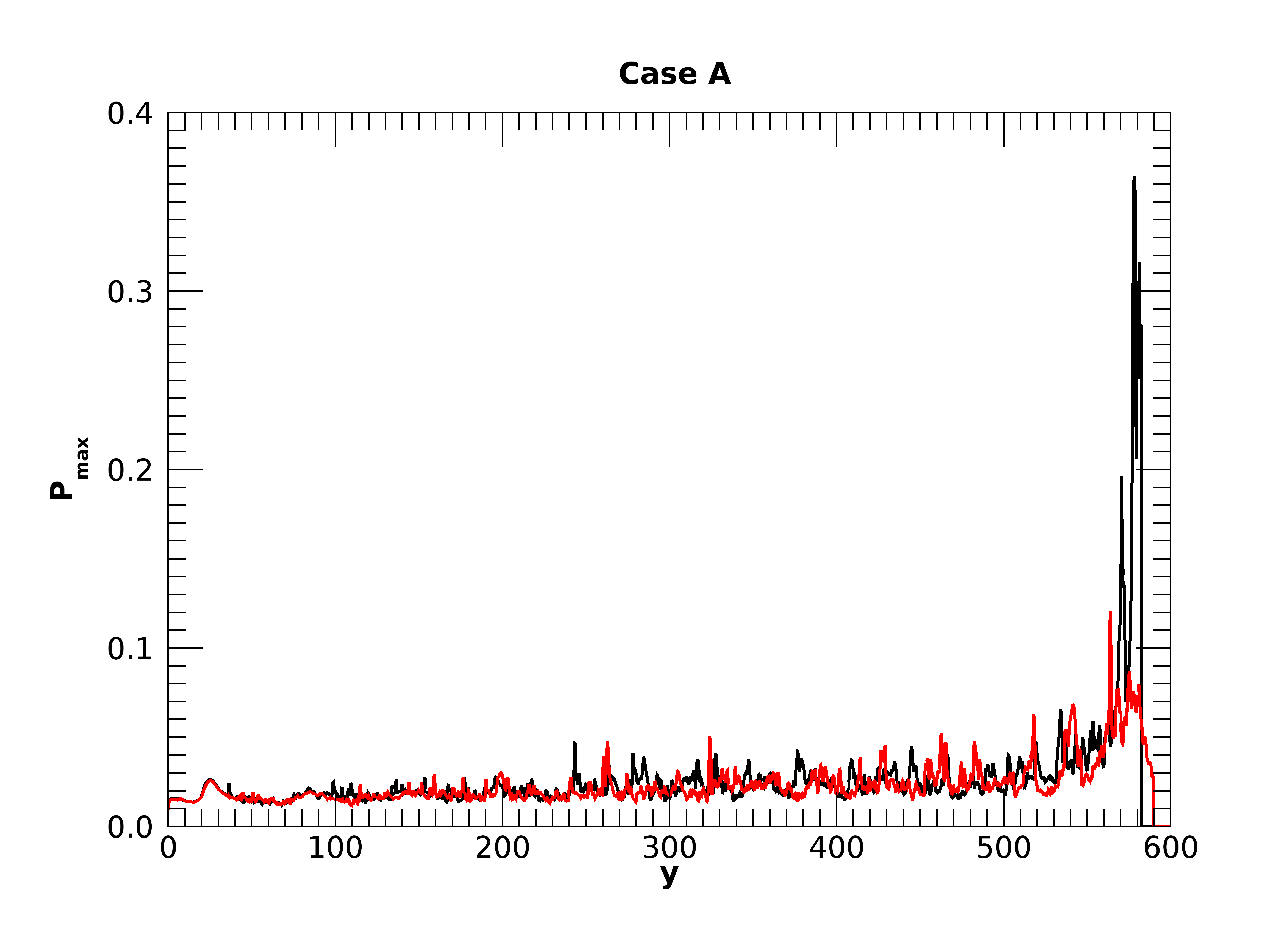}%
\includegraphics[width=6cm]{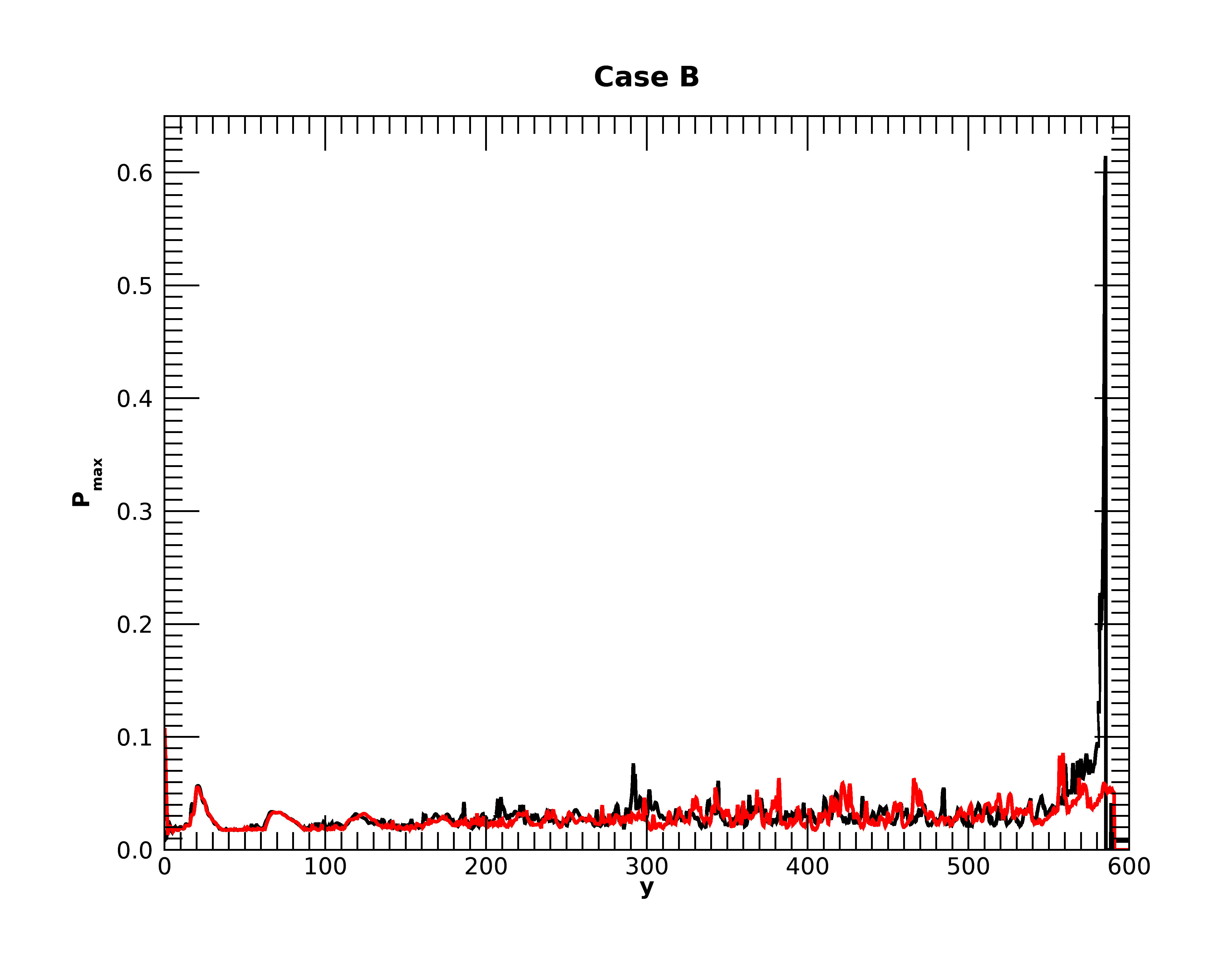}%
\includegraphics[width=6cm]{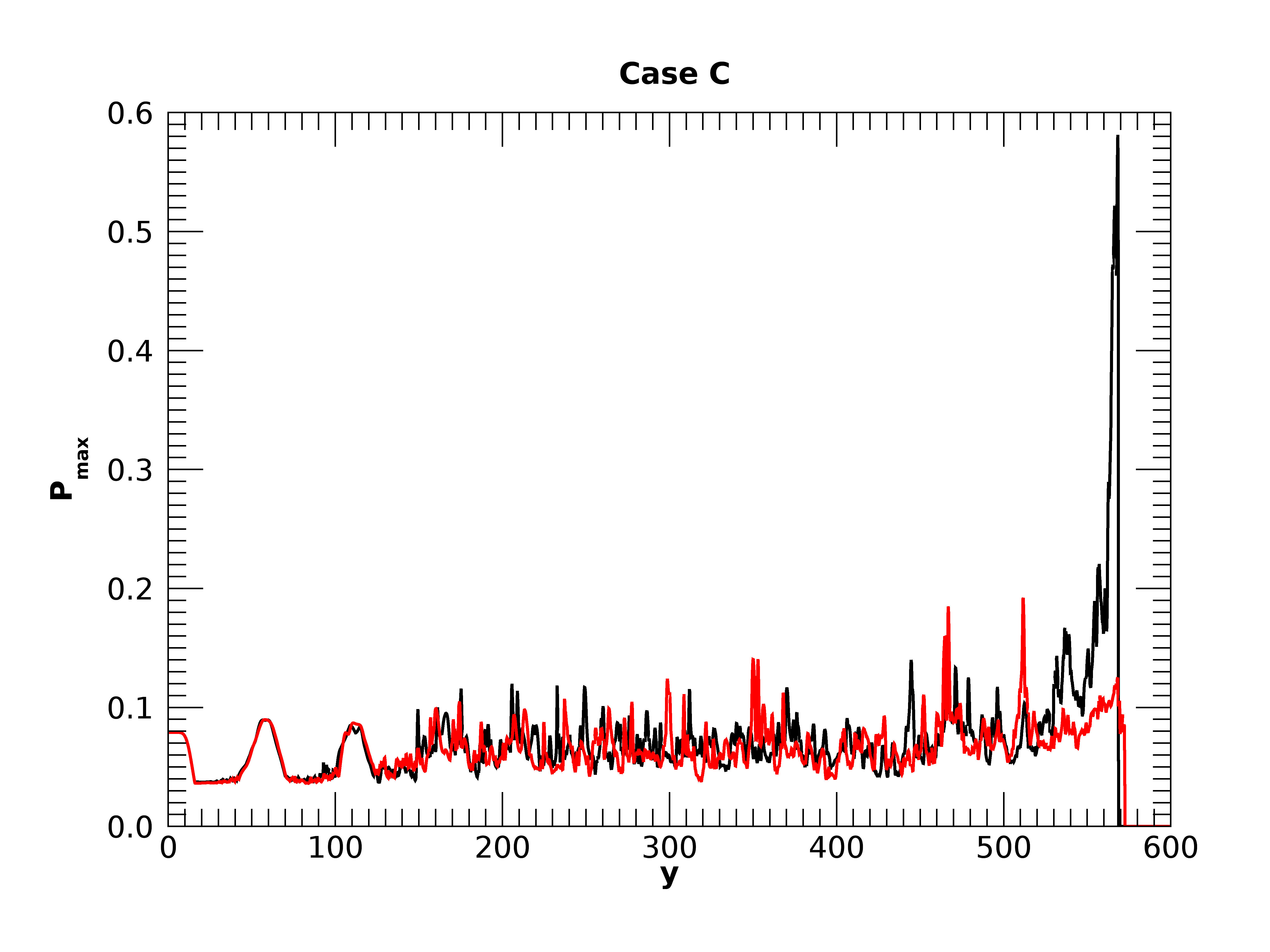}%
\caption{Plots of the maximum pressure on a transverse $x-z$ plane at a given position $y$ along the jet as a function of $y$. The red and the black curves in each panel refer to two different times, but very close, when the jet reaches its maximum extension.
}
\label{fig:pmax} 
\end{figure*}

\begin{figure*}[!htb] 
\centering%
\includegraphics[width=1.9\columnwidth]{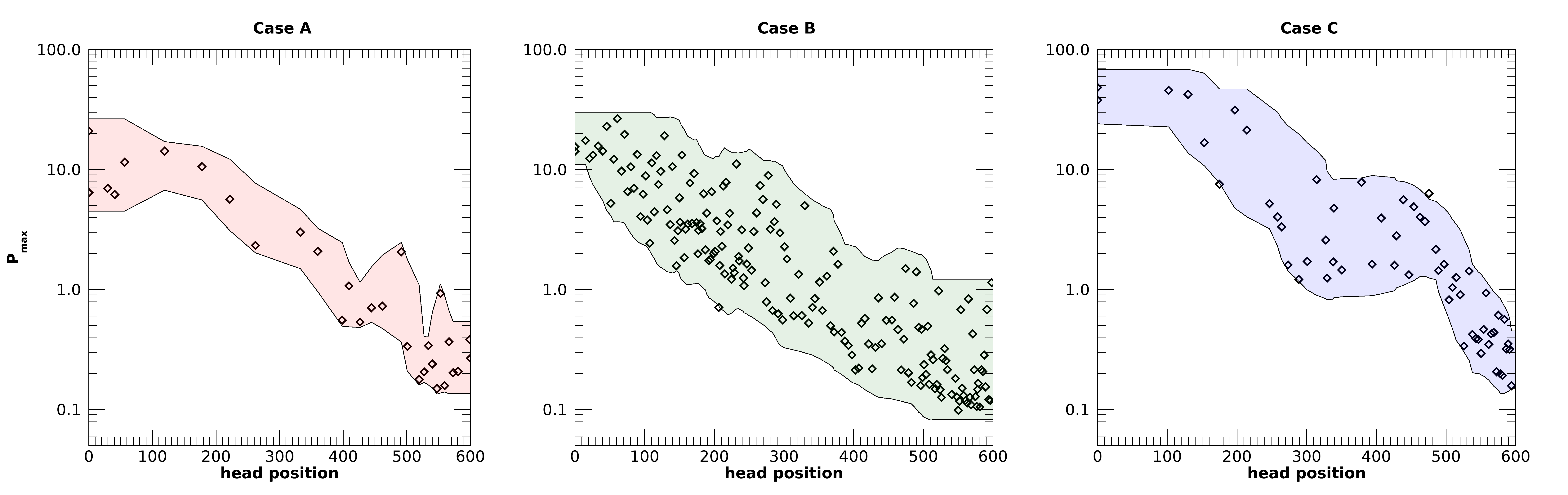}%
\caption{Plots of the maximum pressure found in the computational domain as a function of the position reached by the jet head, the symbols mark the value for each time, while the colored band identifies the range of variation.  The three cases have different numbers of point values, due to the different jet head velocity, a lower velocity (case B) implies a larger number of marks.
}
\label{fig:fig10.png} 
\end{figure*}

\begin{figure}[!ht] 
\centering%
\includegraphics[width=1\columnwidth]{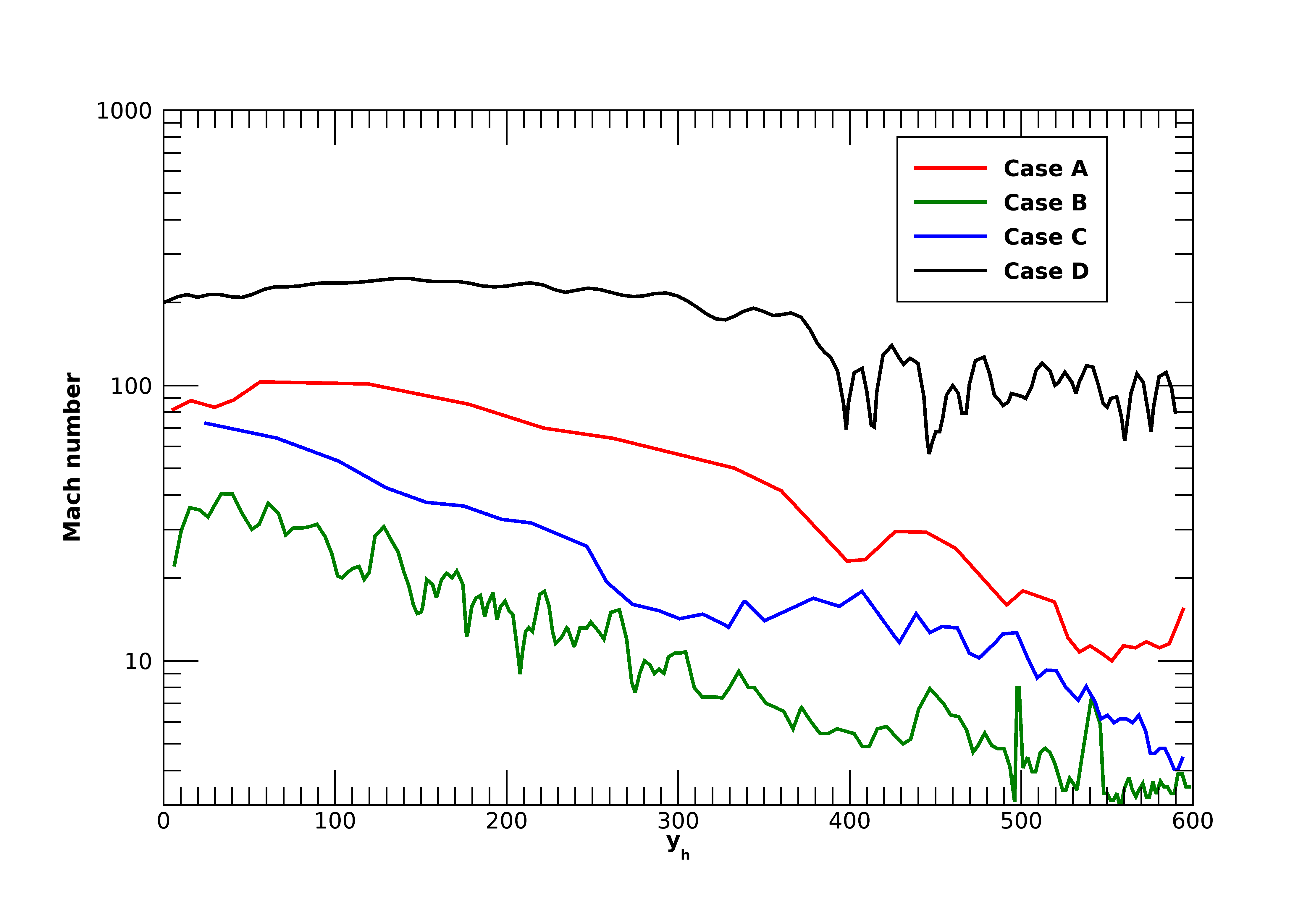}%
\caption{Plots  of jet head Mach number, relative to the ambient medium, as a function of the head position, the three curves refer to the three cases.
}
\label{fig:fig9.png} 
\end{figure}

\begin{figure*}[!ht] 
\centering%
\includegraphics[width=1\textwidth]{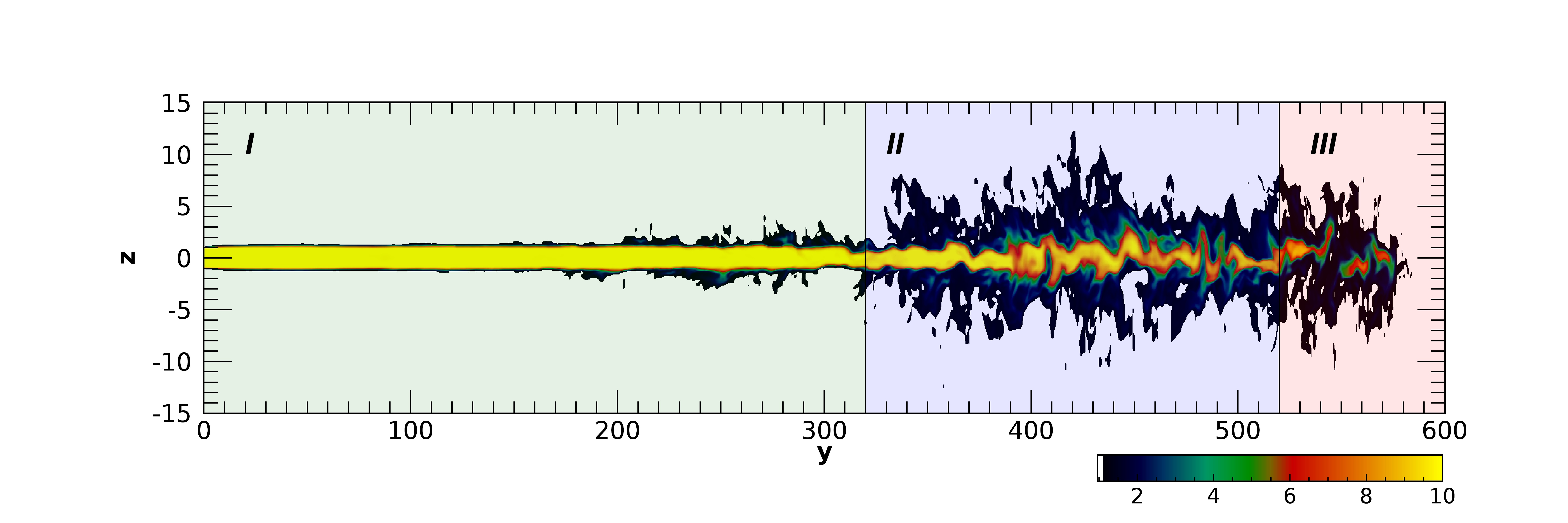}%
\caption{Longitudinal cut  (in the $y-z$ plane)  of the distribution of the Lorentz factor for case D at $t=1,300$ when the jet head reaches $y=600$. The three colored bands highlight the three phases described in the text.
}
\label{fig:gammacut_caseD} 
\end{figure*}

\begin{figure}[!ht] 
\centering%
\includegraphics[width=1\columnwidth]{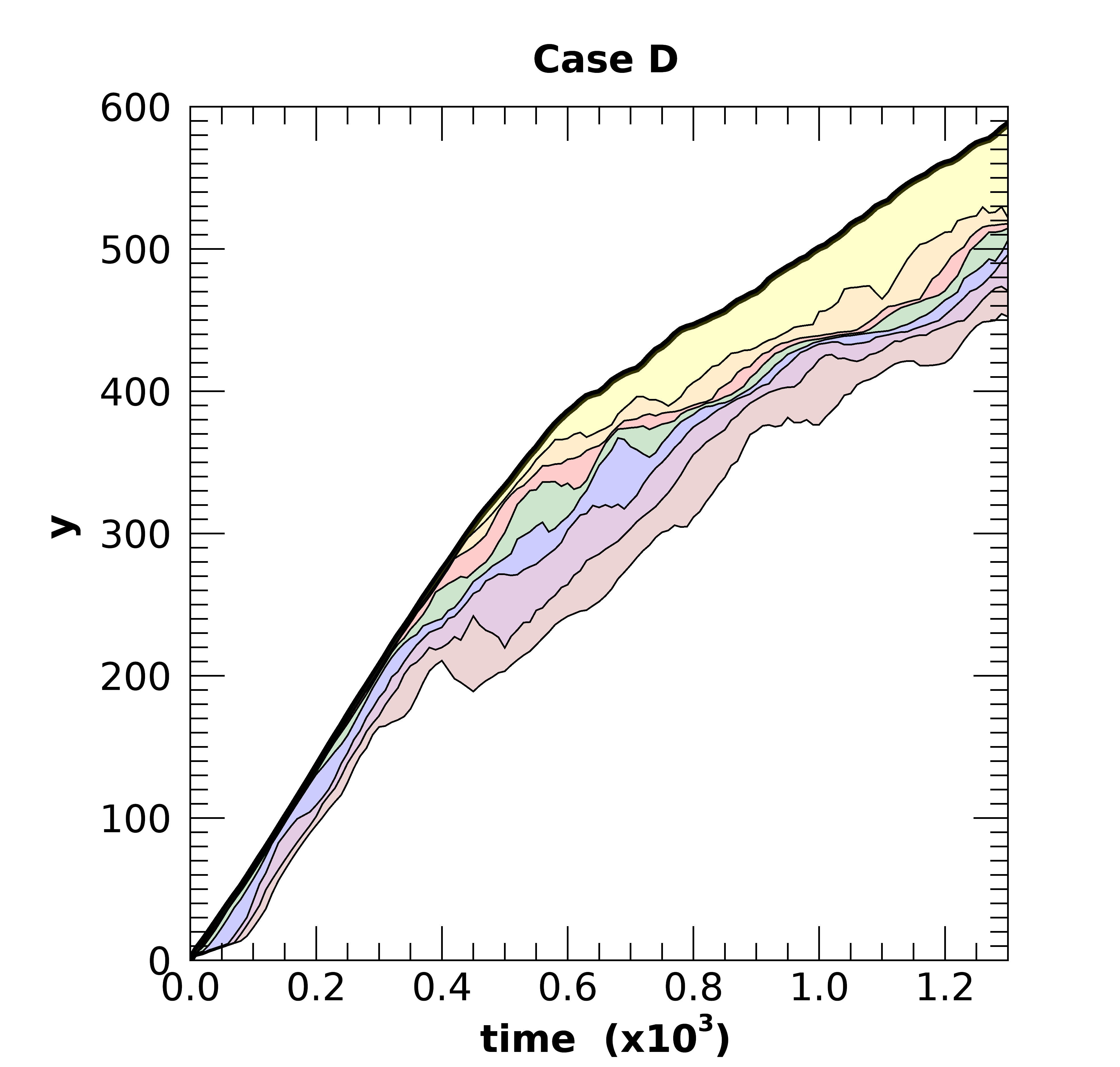}%
\caption{ The figure shows a plot of the maximum distance from the jet origin at which a given fraction of momentum flux is carried by material moving at $\gamma \geq 5$ as a function of time.
From the top we show: the jet head position, and the curves from $10\%$ to $70\%$ differing by $10\%$. We highlight the areas between to subsequent curves with different colors.
}
\label{fig:fig12.png} 
\end{figure}

In Fig. \ref{fig:figmomfr} we give, for case B,  a different view of the  same distribution shown in Fig. \ref{fig:fig7.png}. We plot two vertical cuts at two different times of middle panel of Fig. \ref{fig:fig7.png}, that the fraction of momentum flux carried by high velocity material ($\gamma \geq 5$) as a function of $y$. The curve at later time (red curve) is shifted by few tens of length units, while the jet head, in the same time interval, advanced from $y\sim 400$ to $y\sim 600$, indicating the quasi steady state reached by the jet, as previously discussed. Furthermore, as stressed before, the high velocity material appears to be fragmented in the front part of the jet.

Combining the information obtained from Figs. \ref{fig:r_max}, \ref{fig:fig7.png} and \ref{fig:figmomfr}, we derive a scenario in which the jet, as a result of instabilities, progressively, in a quasi steady way  releases  its momentum to the surrounding material, widens its cross section maintaining a high gamma value only in a central spine until it breaks and fragments into high velocity blobs, surrounded  by an envelop at $\gamma \leq 2$.   We can see that all the three cases  up to $y\approx 400-500$ show a quasi steady state, however they differ in the fraction of momentum flux lost by the high velocity material, that is about $70\%$ for case A and about $90\%$ for cases B and C.
It may be interesting, at this point, to discuss the differences between cases B and C. At the end of the simulations, when the head has reached  $y=600$, they present a similar distribution of the transfer of momentum flux, however at earlier times the evolution is quite different. In case B the momentum curves flatten quite early, while in case C they show an irregular behavior, reach a maximum and then move back towards the jet origin. The two cases have the same density ratio and differ in the pressure ratio, with case C being overpressured. Case C then carries  higher energy and momentum fluxes. Moreover, for all cases, in the first part of the simulation, the cocoon is overpressured with respect to the jet and drives recollimation shocks into the jet, which are stronger in cases B and C. Then, as the cocoon expands its pressure decreases and, for case C, becomes of the order of the jet pressure, the recollimation shocks become then  weak or disappear. Finally, when the cocoon pressure becomes much less than the jet pressure, the jet expands and recollimates, again in a series of shocks. This behavior appears to be correlated with the behavior of the curves of the transfer of momentum flux, that reflect the position at which the instabilities have a substantial growth. This connection between recollimation shocks and the starting of instabilities could be related to the recent results by \citet{komissarov18, gourgouliatos18}, who discuss a centrifugally driven instability of such structures, however higher resolution simulations would be needed in order to make a more precise statement.

The spreading of momentum discussed above leads to a decrease of the strength of the terminal shock up to its disappearance, in Mas16   the authors interpret this disappearance as a signature of the transition from FR II to FR I, in fact the presence of a strong terminal shock can be connected with the presence of hot spots typical of FR II radio sources.  In Fig. \ref{fig:pmax} we plot the maximum pressure on a transverse $x-z$ plane at a given position $y$ along the jet, as a function of $y$. The two curves in each panel refer to two different, but very close, times, when the jet has reached its maximum extension. The peaks close to jet origin are related to the recollimation shocks, that, as we previously pointed out, are progressively dissolved by the instability, they are more evident in cases B and C.  In all the cases, the black curve presents a strong terminal shock, that is not present in the red curve. This demonstrates the highly inhomogeneous jet structure discussed above: the strong shock is related to one of the high velocity blob reaching the head.

Fig. \ref{fig:fig10.png} represents the maximum pressure in the computational domain, that in general corresponds to the pressure of the terminal shock, as a function of the head position.  In this and next figures, we choose to do the plot as a function of the jet head position instead of time, to make easier the comparison among the cases. The figure shows clearly that this quantity is highly fluctuating, for this reason we highlighted the variability range with colored bands. All the three cases show a general decrease of the maximum pressure at the terminal shock,  that can be correlated with a deceleration of the head velocity towards low Mach numbers.
This deceleration is confirmed by Fig. \ref{fig:fig9.png} in  which we plot  the  head Mach number, relative to the external sound speed, as a function of head position. We note how this number approaches  unity ($M=3-4$) for cases B and C, but remains larger for case A ($M=10-20$). Moreover while the curve for case A is smooth and start to present oscillations only in the last part, the wiggling portion widens for case C and, for case B, the head velocity is very variable from the beginning. This indicates that the  fragmentation of the jet occurs earlier in time for case B and progressively later for cases C and A, and may be related to a stronger entrainment and mixing for case B in comparison to cases C and A. Indeed, the width of the variation of the maximum pressure plotted in Fig. \ref{fig:fig10.png} is larger for case B. In the same figure we also plot the head Mach number for case D (black curve). For this last case we see very little deceleration: the Mach number decreases from $\sim 200$ to slightly less than $100$. 

Now, we discuss separately case D, that has a higher density ratio ($\eta = 10^{-2}$) and therefore a larger jet power. On the basis of the results presented so far, we can expect a much less efficient deceleration and indeed  Fig. \ref{fig:fig9.png} confirms this expectation. A weaker interaction with the ambient medium is demonstrated by Fig. \ref{fig:gammacut_caseD}, where we show a longitudinal  cut (in the $y-z$ plane) of the distribution of the $\gamma$ factor. As we have done previously we identify three separate regions: the first one, in which the jet propagates straight, extends up to $y\sim 320$, much beyond the other cases; the second one, in which the instability leads to a jet deformation, covers almost completely the remaining portion of the jet.  The third one is practically absent, and in fact the jet fragmentation occurs only at its head. As a result we see that material moving at $\gamma > 8$ reaches almost continuously the jet front. A further confirmation comes from Fig. \ref{fig:fig12.png},  where similarly to Fig. \ref{fig:fig7.png} we plot, as a function of time the maximum distance at which a given fraction of the momentum flux is carried by material moving at $\gamma \geq 5$. Differently  from the other cases presented in Fig. \ref{fig:fig7.png} all the curves grow with the same slope as the one related to the jet head position, indicating that no steady state has been reached and all the momentum transfer happens always at the jet head.
We can conclude that heavier and more powerful jets cannot be efficiently decelerated within the galaxy core region.

\begin{figure}[!ht] 
\centering%
\includegraphics[width=\columnwidth]{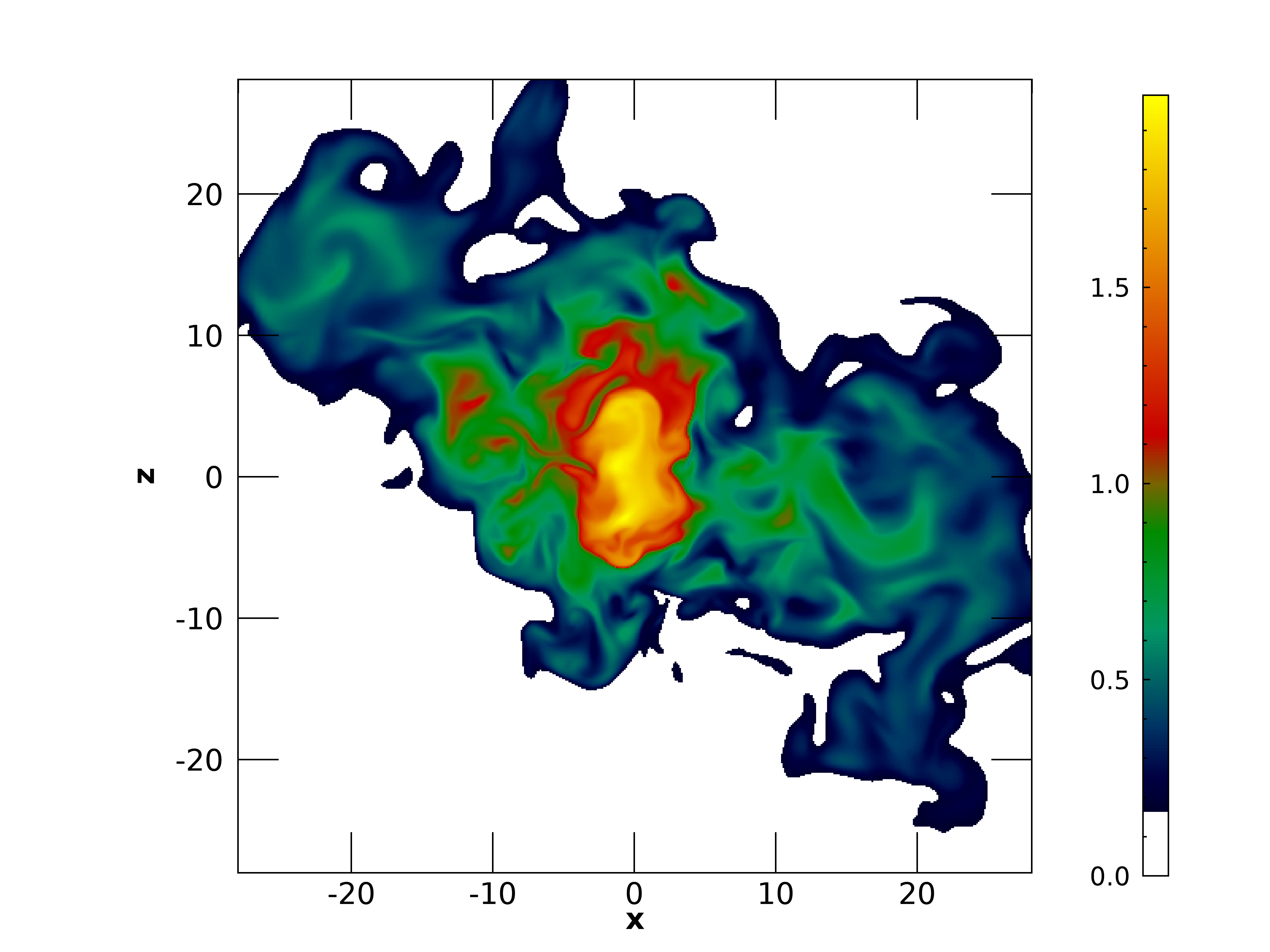}%

\caption{Transverse  cut (in the $x-z$ plane) of the Mach number for case B at $y=450$ and $t=25000$.
}
\label{fig:injection} 
\end{figure}

\begin{figure}[!ht] 
\centering%
\includegraphics[width=\columnwidth]{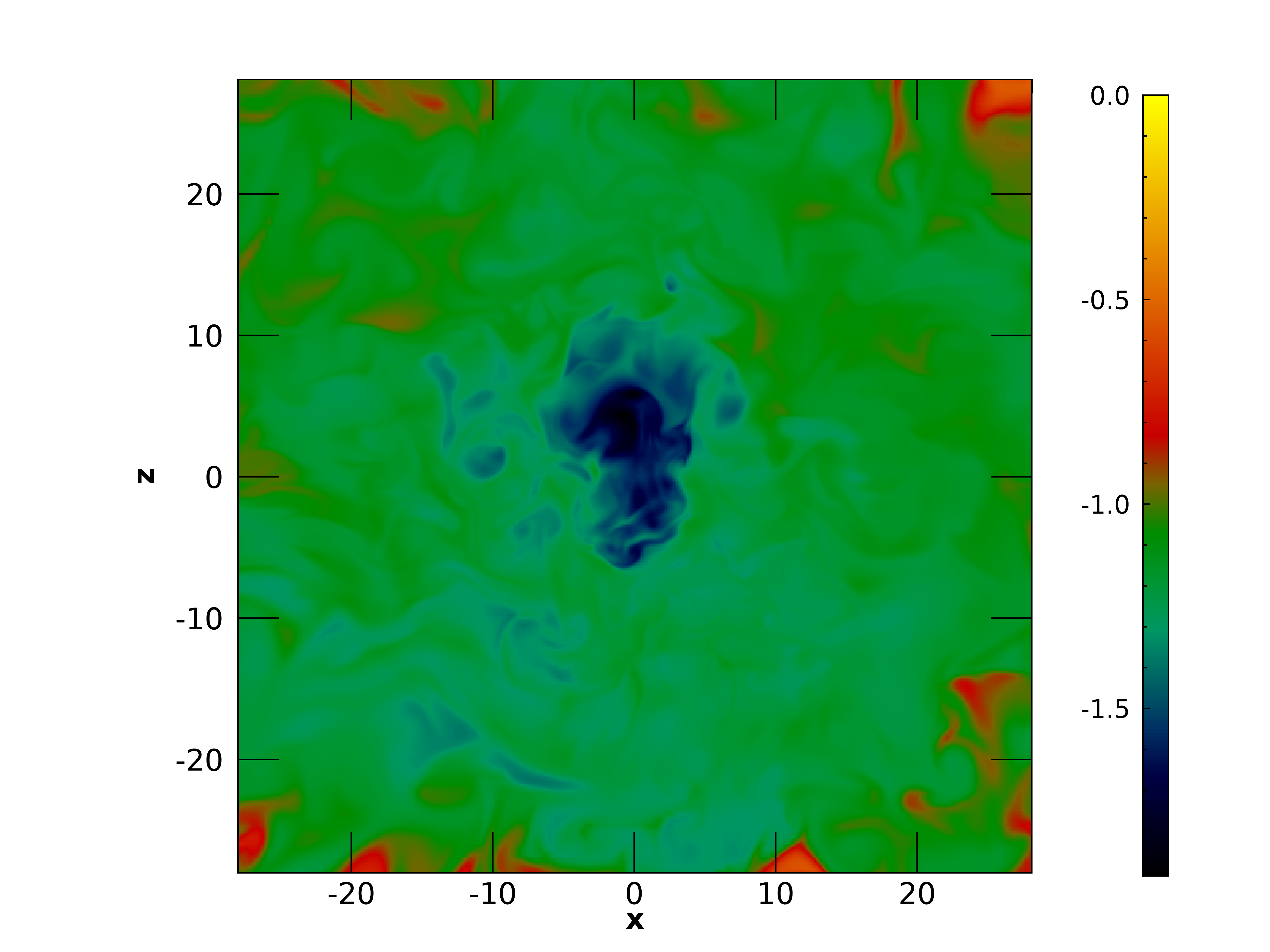}%

\caption{Transverse  cut (in the $x-z$ plane) of the density for case B at $y=450$ and $t=25000$.
}
\label{fig:injdens} 
\end{figure}
 As an aside, we remark that our results confirm that the assumptions of Mas16  are roughly consistent with the results achieved in this paper. In Figs. \ref{fig:injection} and \ref{fig:injdens} we show a typical distribution of the Mach number and of the density in the terminal part of the jet. The maximum Mach number is $\sim 2$, with a corresponding density of $\sim 10^{-2}$, while in the surrounding region the density is $\sim 10^{-1}$. This demonstrates a strong mixing with the external medium, considering that the injected jet density was $10^{-4}$. We notice that our results  are  consistent with the model by \citet{Bicknell95} who estimated that flows decelerating to sub-relativistic velocities acquire a Mach number of the order of 2.

\section{Discussion and conclusions}
\label{sec:discussion}

We presented 3D simulations of relativistic jets, considering four cases with the same Lorentz factor $\gamma=10$ but with different density and pressure ratios and, correspondingly, different powers. These results can be interpreted in relation  to the FR I--FR II classification, assuming a jet injection radius of the order of  one parsec, the powers span an interval between the two classes. Case D can be taken as a reference case, being clearly on the FR II side.  We follow the propagation up to $600$ pc, within the galaxy core, so our assumption of constant density is well justified. In Mas16 , by performing Newtonian simulations, the authors showed that low power jets give rise to a turbulent morphology typical of FRI radio-sources, and give constraints on the physical parameters at the injection point for obtaining such morphology. In this respect, our simulations aim at answering these questions: we know that FR I jets are relativistic at their base, can they decelerate to sub-relativistic  velocity within the first kiloparsec? Are the jet parameters obtained at the distance of about one kiloparsec comparable to the injection parameters assumed in Mas16?

Our results show that relativistic jet deceleration occurs as the consequence of 
 instabilities, which in their growth progressively deform the jet, induce mixing and transfer of momentum to the ambient medium,  and finally lead to a fragmentation of the jet itself.  However another important question arises: the deceleration process described above occurs in a quasi steady way or the jet is capable of reestablishing itself and the deceleration length increases as the jet pierces its way through the ambient medium? We showed evidences that indeed, during the time span of our simulations, a quasi-steady state is reached at least up to $\sim 500$ pc. Nevertheless, cases A, B and C differ in the fraction of momentum transferred (by high velocity material) to the ambient: $90\%$ for the cases B and C, of the order of $70\%$ for the case A. In all the three cases the resulting jet structure, at the end of the steady state region, is a broad flow, moving at about $\gamma \leq 2$, interspersed  by blobs at $\gamma \geq 5$. Extrapolating these results, if we increase $\eta$, and consequently the jet power, the length of the steady deceleration region will likely increase and the transferred fraction of momentum will decrease, indeed case D with $\eta =10^{-2}$ does not show any deceleration effect. Then, if the jet does not decelerate inside the galaxy core it is unlikely that the deceleration will happen in the intergalactic medium with its declining density. On the other hand, decreasing $\eta$, we expect an increase of the transferred momentum fraction with a consequent stronger deceleration. In this respect we can speculate that jets with very low $\eta$ could be not able to propagate much farther than the galaxy core, and could be connected to FR 0 radio-sources. The cases A, B and C seem to represent limiting cases for which an effective deceleration may occur inside the galaxy core.  
 The important role played by the density ratio, in the instability evolution and in the mixing/entrainment process, has been previously investigated both considering jet propagation and more idealized settings \citep[see, e.g.,][]{bodo95,perucho04,perucho05,rossi08} and our present results confirm their findings. The role of instabilities on the jet dynamics, at much larger scales, has  been recently investigated by \citet{perucho19} and \citet{mukherjee20}, who showed that instabilities may lead to some deceleration at larger scales also in higher power jets.

 Furthermore, we can confirm that the assumptions of Mas16   regarding the jet injection parameters are roughly consistent with the results achieved in this paper, as shown in Figs. \ref{fig:injection} and \ref{fig:injdens}  that display typical distributions of the Mach number and of the density in the terminal part of the jet. Clearly the assumption, used in Mas16 , of a spatially and temporally homogeneous flow is very simplified, our results show that there are intervals of time in which high velocity confined structures may still be present; this behavior could provide a perspective on further investigations, along the line of Mas16, in which  these inhomogeneities  can be taken into account.

 Our simulations consider a homogeneous external medium, the presence of strong  inhomogeneities could have an impact on the jet dynamics, as shown by \citet{mukherjee16}, possibly enhancing the entrainment effect. Finally the presence of magnetic field introduces new kinds of instabilities whose importance depends on the field strength and configuration \citep{bodo13}. A strong toroidal component could lead on one hand to a stabilization of velocity shear instabilities, that play a major role in the present case, but introduce current driven instabilities that also may lead to a jet deceleration \citep{Mig2010,Tchekhovskoy16,massaglia19,mukherjee20}. On the other hand, with a weaker more longitudinal field velocity shear instabilities may still be the dominant ones.

\begin{acknowledgements}
 We thank Manel Perucho for the helpfull comments.
 We acknowledge  support  by the {\it{Accordo Quadro INAF-CINECA 2017}} for the availability of high performance computing resources. We acknowledge also support from {\it{PRIN MIUR 2015}} (grant number 2015L5EE2Y). Calculations have been carried out at the CINECA and at the  Competence Centre for Scientific Computing (C3S) of the University of Torino. 
 \end{acknowledgements}


\label{lastpage}
\end{document}